\documentclass[nonacm, sigconf]{acmart}    

\pdfoutput=1 

\usepackage{siunitx}
\usepackage{amsmath}
\usepackage{mathtools}
\usepackage{booktabs}   
\usepackage{subcaption}
\usepackage{graphicx}
\usepackage{multirow}
\usepackage[acronym]{glossaries}
\usepackage{tikz}
\usepackage{xfrac}
\usepackage{xcolor}
\graphicspath{ {./figs/} }
\glsdisablehyper

\mathchardef\mhyphen="2D
\newcommand{\matr}[1]{\mathrm{\mathbf{#1}}}

\newif\ifSuppressMemo
\ifSuppressMemo
\newcommand{\memo}[1]{}
\else
\usepackage{color}
\newcommand{\memo}[1]{{\bf \textcolor{red}{[#1]}}}
\fi

\AtBeginDocument{%
  \providecommand\BibTeX{{%
    \normalfont B\kern-0.5em{\scshape i\kern-0.25em b}\kern-0.8em\TeX}}}

\setcopyright{acmcopyright}
\copyrightyear{2020}
\acmYear{2020}
\acmDOI{}
\acmISBN{}
\acmConference[ACMMM '20]{ACMMM '20: ACM Multimedia Conference (ACMMM 20)}{October 12--16, 2020}{Seattle, United States}

\newacronym{m2p}{M2P}{motion-to-photon}
\newacronym{3dof}{3DoF}{three degrees of freedom}
\newacronym{6dof}{6DoF}{six degrees of freedom}
\newacronym{ran}{RAN}{Radio Access Network}
\newacronym{vr}{VR}{virtual reality}
\newacronym{ar}{AR}{augmented reality}
\newacronym{mr}{MR}{mixed reality}
\newacronym{xr}{XR}{extended reality}
\newacronym{hmd}{HMD}{head-mounted display}
\newacronym{mec}{MEC}{Mobile Edge Computing}
\newacronym{lod}{LOD}{level-of-detail}
\newacronym{fov}{FoV}{field-of-view}
\newacronym{qoe}{QoE}{Quality of Experience}
\newacronym{eeg}{EEG}{electroencephalogram}
\newacronym{emg}{EMG}{electromyography}
\newacronym{vp}{VP}{viewport}
\newacronym{sdp}{SDP}{Session Description Protocol}
\newacronym{ice}{ICE}{Interactive Connectivity Establishment}
\newacronym{ws}{WS}{WebSocket}
\newacronym{p2p}{P2P}{peer-to-peer}
\newacronym{slam}{SLAM}{Simultaneous Localization and Mapping}
\newacronym{slerp}{SLERP}{Spherical Linear Interpolation of Rotations}
\newacronym{rtt}{RTT}{round-trip time}
\newacronym{imu}{IMU}{inertial measurement unit}
\newacronym{mae}{MAE}{mean absolute error}
\newacronym{fps}{FPS}{frames per second}
\newacronym{rd}{RD}{rate-distortion}
\newacronym{autoreg}{AutoReg}{autoregressive}
\newacronym{aic}{AIC}{Akaike Information Criterion}
\newacronym{lat}{LAT}{look-ahead time}
\newacronym{dash}{DASH}{Dynamic Adaptive Streaming over HTTP}
\newacronym{ukf}{UKF}{unscented Kalman filter}
\newacronym{ekf}{EKF}{extended Kalman filter}
\newacronym{cdf}{CDF}{cumulative distribution function}

\begin{document}

\title{Kalman Filter-based Head Motion Prediction for Cloud-based Mixed Reality}

\author{Serhan G{\"u}l}
\affiliation{%
\institution{Fraunhofer HHI}
\city{Berlin}
\country{Germany}}
\email{serhan.guel@hhi.fraunhofer.de}

\author{Sebastian Bosse}
\affiliation{%
	\institution{Fraunhofer HHI}
	\city{Berlin}
	\country{Germany}}
\email{sebastian.bosse@hhi.fraunhofer.de}

\author{Dimitri Podborski}
\affiliation{%
\institution{Fraunhofer HHI}
\city{Berlin}
\country{Germany}}
\email{dimitri.podborski@hhi.fraunhofer.de}

\author{Thomas Schierl}
\affiliation{%
\institution{Fraunhofer HHI}
\city{Berlin}
\country{Germany}}
\email{thomas.schierl@hhi.fraunhofer.de}

\author{Cornelius Hellge}
\affiliation{%
\institution{Fraunhofer HHI}
\city{Berlin}
\country{Germany}}
\email{cornelius.hellge@hhi.fraunhofer.de}

\renewcommand{\shortauthors}{G{\"u}l and Bosse, et al.}

\begin{abstract}	
Volumetric video allows viewers to experience highly-realistic 3D content with six degrees of freedom in mixed reality (MR) environments.
Rendering complex volumetric videos can require a prohibitively high amount of computational power for mobile devices.
A promising technique to reduce the computational burden on mobile devices is to perform the rendering at a cloud server.
However, cloud-based rendering systems suffer from an increased interaction (motion-to-photon) latency that may cause registration errors in MR environments.
One way of reducing the effective latency is to predict the viewer's head pose and render the corresponding view from the volumetric video in advance.

In this paper, we design a Kalman filter for head motion prediction in our cloud-based volumetric video streaming system.
We analyze the performance of our approach using recorded head motion traces and compare its performance to an autoregression model for different prediction intervals (look-ahead times).
Our results show that the Kalman filter can predict head orientations \SI{0.5}{degrees} more accurately than the autoregression model for a look-ahead time of \SI{60}{ms}.
\end{abstract}


\maketitle

\section{Introduction}
\label{sec:intro}
With the advances in volumetric capture technologies, volumetric video has been gaining importance for the immersive representation of 3D scenes and objects for \gls{vr} and \gls{ar} applications~\cite{schreer2019icip}.
Combined with highly accurate positional tracking technologies, volumetric video allows users to freely explore \gls{6dof} content and enables novel \gls{mr} applications where highly realistic virtual objects can be placed inside real environments and animated based on user interaction~\cite{eisert2020}.

Geometry of volumetric objects is usually represented using meshes or point clouds.
High-quality volumetric meshes typically contain thousands of polygons, and high-quality point clouds may contain millions to billions of points~\cite{schreer2019, schwarz2019}.
Therefore, rendering complex volumetric content is still a very demanding task despite the remarkable computing power available in today's mobile devices~\cite{clemm2020}.
Moreover, no efficient hardware implementations of mesh/point cloud decoders are available yet. 
Software-based decoding can be prohibitively expensive in terms of battery usage and may not be able to meet the real-time rendering requirements~\cite{qian2019}.

One way to avoid the complex rendering on mobile devices is to offload the processing to a powerful remote server which dynamically renders a 2D view from the volumetric video based on the user's actual head pose~\cite{shi2015}. 
The server then compresses the rendered texture into a 2D video stream and transmits it over a network to the client.
The client can then efficiently decode the video stream using its hardware decoder and display the dynamically updated content to the viewer.
Moreover, the cloud-based rendering approach allows utilizing highly efficient 2D video coding techniques and thus can reduce the network bandwidth requirements by avoiding the transmission of the volumetric content~\cite{qian2019}.

However, one drawback of cloud-based rendering is the increased interaction latency, also known as the \gls{m2p} latency~\cite{shi2012}. 
Due to the network round-trip time and the added processing delays, the \gls{m2p} latency is higher than in a system that performs the rendering locally. 
Several studies show that an increased interaction latency may lead to a degraded user experience and motion sickness~\cite{beigbeder2004, mccandless2000, livingston2008}.

One way to reduce the latency is to predict the user's future head pose at the cloud server and render the corresponding view of the volumetric content in advance.
Thereby, it is possible to significantly reduce or even eliminate the \gls{m2p} latency, if the user pose is successfully predicted for a \gls{lat} equal to or larger than the \gls{m2p} latency of the system. 
However, mispredictions of head motion may increase registration errors and degrade the user experience in AR environments~\cite{livingston2008}. 
Therefore, designing accurate head motion prediction algorithms is crucial for high-quality volumetric video streaming.

In this paper, we consider the problem of head motion prediction for cloud-based \gls{ar}/\gls{mr} applications.
Our main contributions are as follows:
\begin{itemize}
	\item We develop a Kalman filter-based predictor for head motion prediction in \gls{6dof} space and analyze its performance compared to an autoregression model and a baseline (no prediction) model using recorded head motion traces.
	\item We present an architecture for integration of the Kalman filter-based predictor into our existing cloud-based volumetric streaming framework.
\end{itemize}

The paper is structured as follows. 
Sec.~\ref{sec:background} gives an overview of the literature on  volumetric video streaming, remote rendering and head motion prediction.
Sec.~\ref{sec:system} gives an overview of our cloud-based volumetric video streaming system.
Sec.~\ref{sec:prediction} describes the developed Kalman filter-based predictor and presents a framework for its integration into our volumetric streaming system.
Sec.~\ref{sec:evaluation} presents our experimental setup and the evaluation results. 
Sec.~\ref{sec:conclusion} concludes this paper.

\section{Related Work}
\label{sec:background}

\subsection{Volumetric video streaming}
\label{sec:volumetric}
A few recent works deal with efficient streaming of volumetric videos in different content representations.
Hosseini and Timmerer \cite{hosseini2018} extended the concepts of \gls{dash} for point cloud streaming. They proposed different approaches for spatial subsampling of dynamic point clouds to decrease the density of points in the 3D space and thus reduce the bandwidth requirements. 
Park et al.~\cite{park2019} proposed using 3D tiles for streaming of voxelized point clouds.
Their system selects 3D tiles and adjusts the corresponding \glspl{lod} using a rate-utility model that considers the user's viewpoint and distance to the object. 
Qian et al.~\cite{qian2019} developed a point cloud streaming system that uses an edge proxy to convert point cloud streams into 2D video streams based on the user's viewpoint in order to enable efficient decoding on mobile devices. 
They also proposed various optimizations to reduce the \gls{m2p} latency between the client and the edge proxy.
Van der Hooft et al.~\cite{vanderHooft2019} proposed an adaptive streaming framework compliant to the recent point cloud compression standard MPEG V-PCC~\cite{schwarz2019}. 
Their framework PCC-DASH enables adaptive streaming of scenes with multiple dynamic point cloud objects.
They also presented rate adaptation techniques that rely on the user's position and focus as well as the available bandwidth and the client's buffer status to select the optimal quality representation for each object.
Petrangeli et al.~\cite{petrangeli2019} proposed a streaming framework for AR applications that dynamically decides which virtual objects should be fetched from the server as well as their \glspl{lod}, depending on the proximity of the user and likelihood of the user to view the object.


\subsection{Remote rendering}
\label{sec:remote}
The idea of offloading the rendering process to a powerful remote server was first considered in 1990s when PCs did not have sufficient computational power for intensive graphics tasks~\cite{shi2015}.
A remote rendering system renders complex graphics on a powerful server and delivers the result over a network to a less-powerful client device.

With the advent of cloud gaming and \gls{mec}, \emph{interactive} remote rendering applications have started to emerge which allow the client device to control the rendering application based on user interaction~\cite{mangiante2017, shi2019, qian2019}.
Mangiante et al.~\cite{mangiante2017} presented a \gls{mec} system for \gls{fov} rendering of \ang{360} videos to optimize the required bandwidth and reduce the processing requirements and battery utilization.
Shi et al.~\cite{shi2019} developed a \gls{mec} system to stream \gls{ar} scenes containing only the user's \gls{fov} plus a latency-adaptive margin around it. 
They deployed the prototype on a \gls{mec} node connected to a LTE network and evaluated its performance.
A detailed survey of interactive remote rendering system is given in~\cite{shi2015}.

\subsection{Head motion prediction}
\label{sec:prediction_bg}
Previous techniques for head motion prediction were mainly developed for dealing with the rendering and display delays of the early AR systems.
In his dissertation, Azuma~\cite{azuma1995} developed an AR system that relies on head motion prediction to reduce the dynamic registration errors of virtual objects. His results indicate that prediction is most effective for short prediction intervals that are less than \SI{80}{ms}.
In a follow-up work, Azuma and Bishop~\cite{azuma1995_freq} presented a frequency domain analysis of head motion prediction and concluded that the error in predicted position grows rapidly with increasing prediction intervals and head motion signal frequencies.
Van Rhijn et al.~\cite{van_rhijn2005} proposed a framework for Bayesian predictive filtering algorithms and studied the effect of the filter parameters on the prediction performance. 
They compared the performances of different prediction methods using both synthetic and experimental data.
La Viola~\cite{laviola2003} presented a comparison of the \gls{ukf} and \gls{ekf} for prediction of head and hand orientation represented with quaternions.
Kraft~\cite{kraft2003} proposed a quaternion-based \gls{ukf} that extends the original \gls{ukf} formulation to address the inherent properties of unit quaternions. The developed filter was applied for prediction of head orientation; however, the evaluation is limited to simulated motion.
Himberg and Motai~\cite{himberg2009} proposed an \gls{ekf} that operates on the change of quaternions between consecutive time points (delta quaternion) and showed that their approach provides similar prediction performance to a quaternion-based \gls{ekf} with less computational burden.

In recent years, with the resurgence of interest in \gls{vr}, head motion prediction regained importance for prediction of the future user viewport in \ang{360} videos.
Bao et al.~\cite{bao2016} developed regression models to predict the user's viewport. 
They also used the same models to predict the accuracy of prediction to determine the size of margins around the viewport for efficient transmission of \ang{360} videos.
Sanchez et al.~\cite{sanchez2019} proposed angular velocity and angular acceleration based predictors to tackle the delay issue in tile-based viewport-dependent streaming.
Qian et al.~\cite{qian2018_flare} analyzed the performance of several machine learning algorithms on head movement traces collected from 130 diverse users and employed different prediction algorithms depending on the prediction interval. 
The developed prediction framework was integrated into a streaming system for \ang{360} videos to reduce the bandwidth usage or boost the video quality given the same bandwidth.

The principles of head motion prediction for \ang{360} videos and volumetric videos in mixed reality are similar. 
However, volumetric videos are more complex and allow movement in a higher degree of freedom making prediction a more difficult task.
In our volumetric streaming system, we employ a Kalman filter-based prediction framework to jointly predict both translational and rotational head movements in \gls{6dof} space.

\section{Volumetric streaming system}
\label{sec:system}
\begin{figure}[t]
	\includegraphics[width=\linewidth]{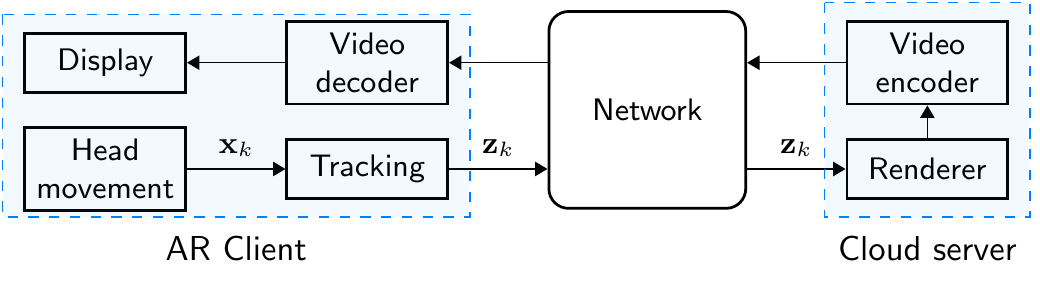}
	\caption[System model]{High level operation of our cloud-based volumetric streaming system.}
	\label{fig:system_baseline}
\end{figure}

Fig.~\ref{fig:system_baseline} shows an overview of our cloud-based volumetric streaming system. 
We abstract the software components as functional blocks to focus on the prediction aspects.
A detailed software architecture of our system\footnote{Reference is hidden for peer review} is described in~\cite{anonym}.

At the cloud server, we store our compressed volumetric video as a single MP4 file containing video and mesh tracks. 
Particularly, we encode the texture atlas using H.264/AVC and the mesh geometry using  Google Draco that implements the Edgebreaker algorithm~\cite{rossignac1999}.
The compressed mesh and texture data are multiplexed into different tracks of an MP4 file ready to be processed by the game engine (Unity) running at our server by means of a native plug-in that demultiplexes and decodes the respective data streams.

Head movement of the user is described by the state vector $\mathbf{x}_k$.
The tracking system of the \gls{ar} headset measures the head pose with a sampling interval of $t_s$. 
The measurement $\mathbf{z}_k$ is then sent over a network to the cloud server, which then renders the corresponding view from the volumetric content (textured meshes) according to $\mathbf{z}_k$.
Next, the rendered view is encoded as a video stream using the NVIDIA hardware encoder (NVENC)~\cite{nvidia2019} and sent to the client using WebRTC~\cite{webrtc}. 
We selected WebRTC as the delivery protocol since it provides low-latency (real-time) streaming capabilities and is already widely adopted by different web browsers allowing our system to support several different platforms.
After the transmission, the client decodes the received video stream and displays it to the viewer.

The time period between the head movement and display of the decoded video frame to the viewer is the \gls{m2p} latency of the system which we aim to compensate by applying prediction.

\section{Head motion prediction}
\label{sec:prediction}

We propose to use a Kalman filter for \gls{6dof} head motion prediction in our cloud-based volumetric streaming system.
As a benchmark, we investigate the performance of a Baseline and an autoregression model.
\subsection{Baseline}
The Baseline model represents the operation of the system without prediction.
We assume that the prediction time is set equal to the \gls{m2p} latency such that the prediction completely eliminates the latency.
For a prediction time of $N$ samples, the measurement $\mathbf{z}_k$ is simply propagated $N$ samples ahead in our simulations and set as the user pose at time $k+N$, i.e. $\hat{\mathbf{x}}_{k+N}=\mathbf{z}_k$.

\subsection{Autoregression}
\label{sec:autoreg}
\Gls{autoreg} models use a linear combination of the past values of a variable to forecast its future values~\cite{hyndman2018}.

An \gls{autoreg}  model of lag order $\rho$ can be written as
\begin{equation}
\label{eq:autoreg}
    y_t = c+\phi_1 y_{t-1}+\phi_2 y_{t-2}+\dots+\phi_\rho y_{\rho-1}+\epsilon_t
\end{equation}
where $y_t$ is the true value of the time series $y$ at time $t$, $\epsilon_t$ is the white noise, $\phi_i$ are the coefficients of the model.
Such a model with $\rho$ lagged values is referred to as an $\mathrm{AR}(\rho)$ model. 
The optimal lag order for the model can be automatically determined using statistical tests such as the \gls{aic}~\cite{akaike1973}.
\gls{autoreg} models must first be trained to learn the model coefficients, and the learned model coefficients $\phi_i$ as well as the lag order $\rho$ may vary depending on the training data.

\begin{figure}[ht]
	\includegraphics[width=\linewidth]{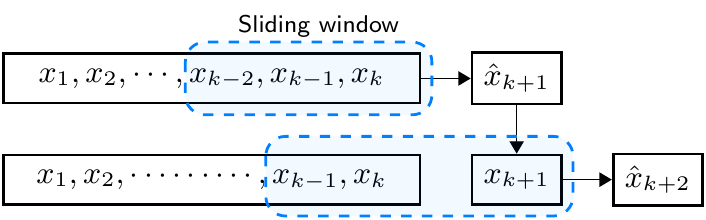}.
	\caption{An example showing multi-step ahead prediction using autoregression model for a history window of 3 and a \acrfull{lat} of 2 samples.}
	\label{fig:autoreg}
\end{figure}

Typically, we need to predict not only the next sample but multiple samples ahead in the future to achieve a given \acrfull{lat}. 
Therefore, we repeat the prediction step in a sliding window fashion by using the just-predicted sample for the prediction of the next sample and iterate Eq.~\eqref{eq:autoreg} until we achieve the desired \gls{lat}.
Fig.~\ref{fig:autoreg} shows an example that demonstrates the iterations of the \gls{autoreg} model for a history window of 3 samples and \gls{lat}=2.

\subsection{Kalman filter}
\subsubsection*{Basics}
\label{sec:kalman}

The Kalman filter estimates the state $\mathbf{x}\in\mathbb{R}^n$ of a
discrete-time process expressed by the linear difference equation
\begin{equation}
\label{eq:discrete_model}
\mathbf{x}_k = \matr{F}\mathbf{x}_{k-1} + \mathbf{w}_{k-1},
\end{equation}
with a measurement (observation) $\mathbf{z}\in\mathbb{R}^m$ expressed by
\begin{equation}
\mathbf{z}_k = \matr{H}\mathbf{x}_k + \mathbf{v}_{k},
\end{equation}
where the random variables $\mathbf{w}_{k}$ and $\mathbf{v}_{k}$ represent the process and measurement noise, respectively. 
They are assumed to be independent from each other and have the Gaussian distributions $\mathit{p}(\mathbf{w})\sim \mathit{N}(0,\matr{Q})$ and $\mathit{p}(\mathbf{v})\sim \mathit{N}(0,\matr{R})$ where $\matr{Q}$ and $\matr{R}$ are the process and measurement noise covariance matrices, respectively.
The general formulation of the Kalman filter allows $\matr{Q}$ and $\matr{R}$ to be changed at each time step; however, we assume that they remain constant.
The matrix $\matr{F}\in\mathbb{R}^{n\times n}$ represents the state transition (process) model that relates the state at the previous time step $\mathbf{x}_{k-1}$ to the current state $\mathbf{x}_{k}$. 
The matrix $\matr{H}\in\mathbb{R}^{m\times n}$ represents the observation model that relates the state to the measurement.
Since no external control is involved in a subject's head movements, we do not include a control input in our framework~\cite{welch1995}.

Given the knowledge of the state before step $k$, we define $\hat{\mathbf{x}}_{k}^\mhyphen$ as our \textit{a priori} state estimate, and given the measurement $\mathbf{z}_k$ at step $k$, we define $\hat{\mathbf{x}}_{k}$ as our \textit{a posteriori} state estimate. 
We can then define \textit{ a priori} estimate error and \textit{a posteriori} estimate error as
\begin{align} 
\matr{P}_k^\mhyphen &= \mathrm{cov}(\mathbf{x}_k-\hat{\mathbf{x}}_{k}^\mhyphen), \\ 
\matr{P}_k &= \mathrm{cov}(\mathbf{x}_k-\hat{\mathbf{x}}_{k}), 
\end{align}
respectively. 

A complete Kalman filter cycle consists of the time update ("prediction") and measurement update ("correction") steps. 
The time update projects the current state estimate ahead in time using the process model $\matr{F}$. 
Given the previous state estimate $\hat{\mathbf{x}}_{k-1}$, the \emph{a priori} state and error covariance estimates are obtained by
\begin{align}
\hat{\mathbf{x}}_{k}^\mhyphen &= \matr{F}\hat{\mathbf{x}}_{k-1}, \label{eq:process}  \\
\matr{P}_{k}^\mhyphen &= \matr{F}\matr{P}_{k-1}\matr{F}^T + \matr{Q},
\end{align}
respectively.

The measurement update applies a correction to the projected estimate using the actual measurement.
First, the Kalman gain $\matr{K}$ is computed which is a ratio expressing how much the filter "trusts" the prediction vs. the measurement
\begin{align}
\matr{K}_k = \matr{P}_{k}^\mhyphen \matr{H}^T(\matr{H}\matr{P}_{k}^\mhyphen \matr{H}^T+\matr{R})^{-1}.
\end{align}
Then, the actual measurement $\mathbf{z}_k$ is incorporated to obtain the \emph{a posteriori} state estimate by 
\begin{align}
\mathbf{\hat{x}}_{k} = \mathbf{\hat{\mathbf{x}}}_{k}^\mhyphen + \matr{K}_k(\mathbf{z}_k-\matr{H}\hat{\mathbf{x}}_{k}^\mhyphen),
\end{align}
and the \emph{a posteriori} error covariance is obtained by
\begin{align}
\matr{P}_{k} = (\matr{I}-\matr{K}_k\matr{H})\matr{P}_{k}^\mhyphen.
\end{align}
A detailed derivation of the Kalman filter equations can be found in~\cite{labbe2015}.

\subsubsection*{Filter design}
\label{sec:filter_design}
We use a 14D state vector that consists of the position and orientation components as well as their first time-derivatives
\begin{equation}
    \mathbf{x} = \left[x,\dot{x},y,\dot{y},z,\dot{z},q_w,\dot{q_w},q_x,\dot{q_x},q_y,\dot{q_y},q_z,\dot{q_z} \right]^T.
\end{equation}
We initialize the state as a zero-vector, $\mathbf{x}=\matr{0}_{14}$ and the error covariance as an identity matrix, $\matr{P}=\matr{I}_{14}$.
Our state transition model $\matr{F}\in\mathbb{R}^{14\times 14}$ is a block diagonal matrix with the block
$$
\begin{bmatrix}
1 & \Delta t \\
0 & 1
\end{bmatrix}
$$
repeated seven times in the diagonal such that the same constant-velocity motion model is applied to all the state variables. 
$\Delta t$ is the time step of the filter set equal to the sampling time $t_s$ (\SI{5}{ms}) in our simulations.
Our observation model $\matr{H}\in\mathbb{R}^{7\times 14}$ is also a block diagonal matrix with $
\begin{bmatrix}
1 & 0 
\end{bmatrix}
$
repeated seven times in the diagonal.

\subsubsection*{Measurement noise}
$\matr{R}\in\mathbb{R}^{7\times 7}$ models the noise in the sensors as a covariance matrix.
The \gls{ar} headset we use for data collection, Microsoft HoloLens, uses advanced visual-inertial \gls{slam} algorithms that track the user pose with high accuracy~\cite{liu2018_hololens}.
Therefore, we experimented with small noise variances and empirically constructed $\matr{R}$ as a diagonal matrix with the diagonal values set to  $10^{-6}$.
In practice, there might be correlation between different sensors and usually their noise is not a pure Gaussian~\cite{labbe2015}.
However, for the lack of a better sensor model, we employ a simplified measurement matrix.

\subsubsection*{Process noise}
Since we have a discrete-time process in which we sample the system at regular intervals, we need a discrete representation of the noise term $\mathbf{w}$ given in Eq.~\ref{eq:discrete_model}.
Therefore, we consider a discretized white noise model which assumes that the velocity remains constant during each $\Delta t$ but differs for each time period~\cite{barshalom2004}.

The process noise covariance matrix $\matr{Q}\in\mathbb{R}^{14\times 14}$ is then a block diagonal matrix with the block
$$
\begin{bmatrix}
\frac{\Delta t^4}{4} & \frac{\Delta t^3}{2} \\[0.5em]
\frac{\Delta t^3}{2} & \Delta t^2
\end{bmatrix}
\sigma_\nu^2
$$
repeated seven times in the diagonal, where the noise variance is empirically set to $\sigma_\nu^2=10^3$ for the first three blocks (position) and $\sigma_\nu^2=4\times10^6$ for the remaining blocks (orientation). 
A derivation of $\matr{Q}$ for the discretized white noise model is given in~\cite{barshalom2004}.

\subsubsection*{Multi-step ahead prediction}
\label{sec:kf_prediction}
Each iteration of the Kalman filter results in an \textit{a posteriori} state estimate $\mathbf{\hat{x}}_{k}$.
To obtain an $N$-step prediction $\mathbf{\hat{x}}_{k+N}$, after each iteration we need to propagate $\mathbf{\hat{x}}_{k}$ ahead by applying the process model $\matr{F}$ multiple times on $\mathbf{\hat{x}}_{k}$, i.e. by iterating the Eq.~\eqref{eq:process} $N$ times~\cite{kiruluta1997, van_rhijn2005}. 

\subsection{Representation of orientation}
\label{sec:orientation}
We perform the prediction of orientations in the quaternion domain~\cite{diebel2006}. 
We readily obtain quaternions from the HoloLens and thus can avoid conversion from another representation such as Euler angles or rotation matrices.
Quaternions allow smooth interpolation of orientations using techniques like \gls{slerp}~\cite{shoemake1985}. 
Moreover, quaternions do not suffer from gimbal lock\footnote{Gimbal lock is the loss of one degree of freedom while using Euler angles, when the pitch angle approaches $\pm \ang{90}$.} as opposed to Euler angles and offer a singularity-free description of orientation.
They are more compact compared to rotation matrices and thus computationally more efficient~\cite{jia2013}. 
The set of unit quaternions, i.e. quaternions of norm one, constitutes a unit sphere in 4-D space. 
The three remaining degrees of freedom after applying the unity constraint are sufficient to represent any rotation in 3-D space~\cite{diebel2006}.

\subsection{System integration}
\label{sec:sys_integration}
\begin{figure}[t]
	\includegraphics[width=\linewidth]{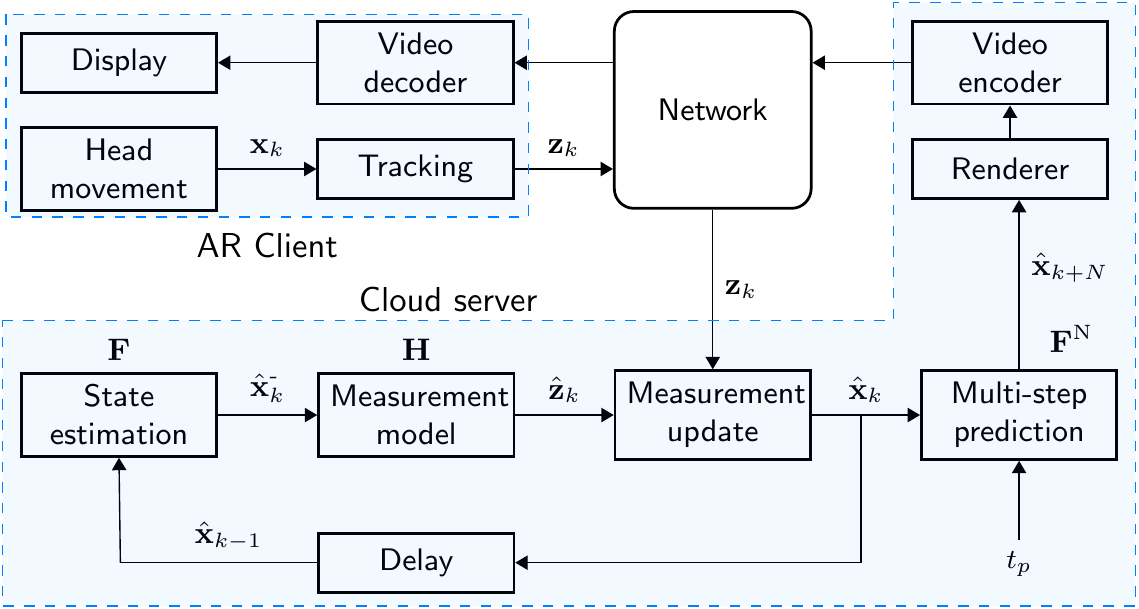}
	\caption[Kalman filter]{Integration of the Kalman filter-based predictor into our cloud-based volumetric streaming system.}
	\label{fig:system_kalman}
\end{figure}
\begin{figure}[t]
	\centering
	\includegraphics[width=\linewidth]{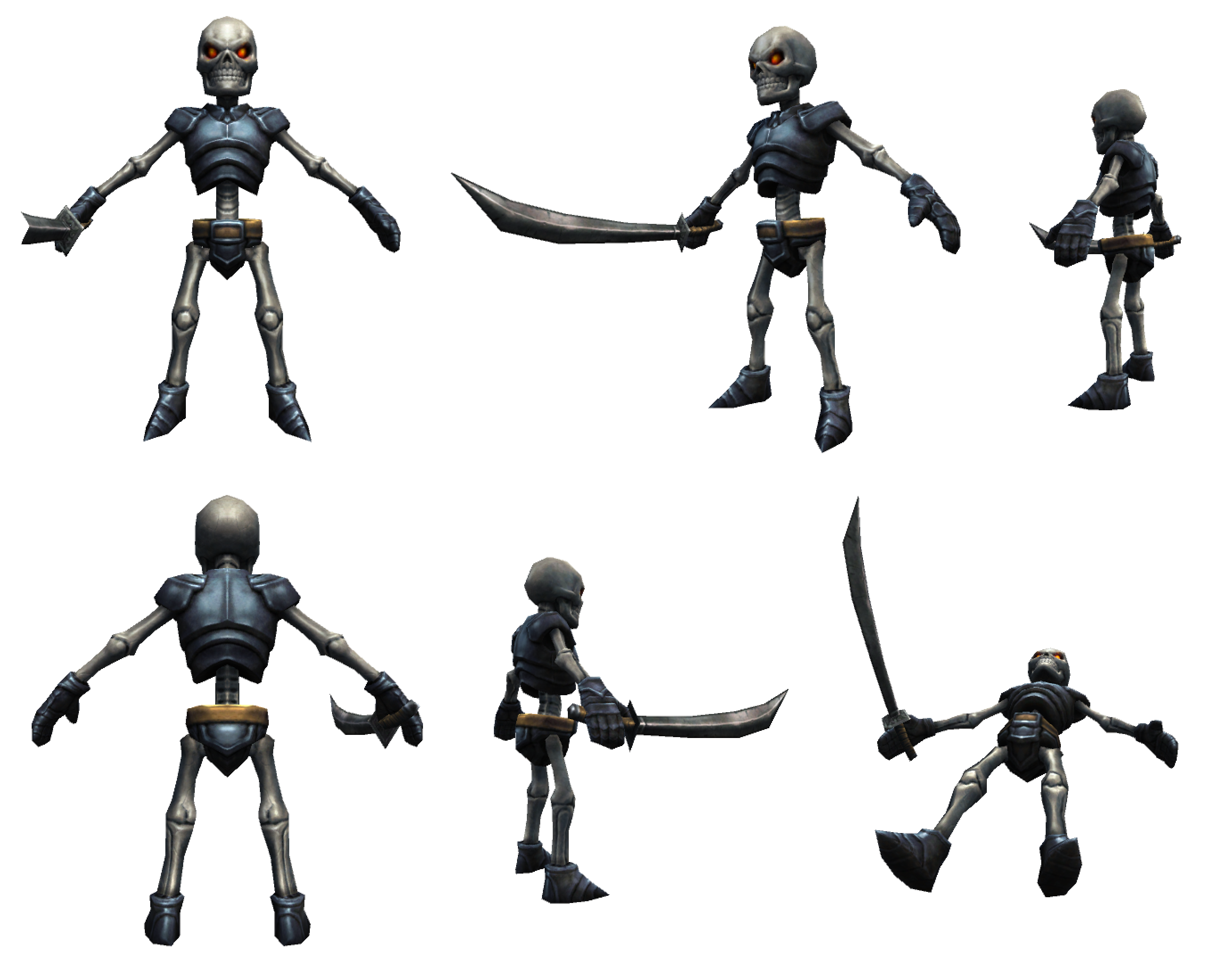}
	\caption{The virtual object used during trace collection shown from different viewpoints.}
	\label{fig:skeleton}
\end{figure}

\begin{figure*}[t]
	\centering
	\includegraphics[width=\linewidth]{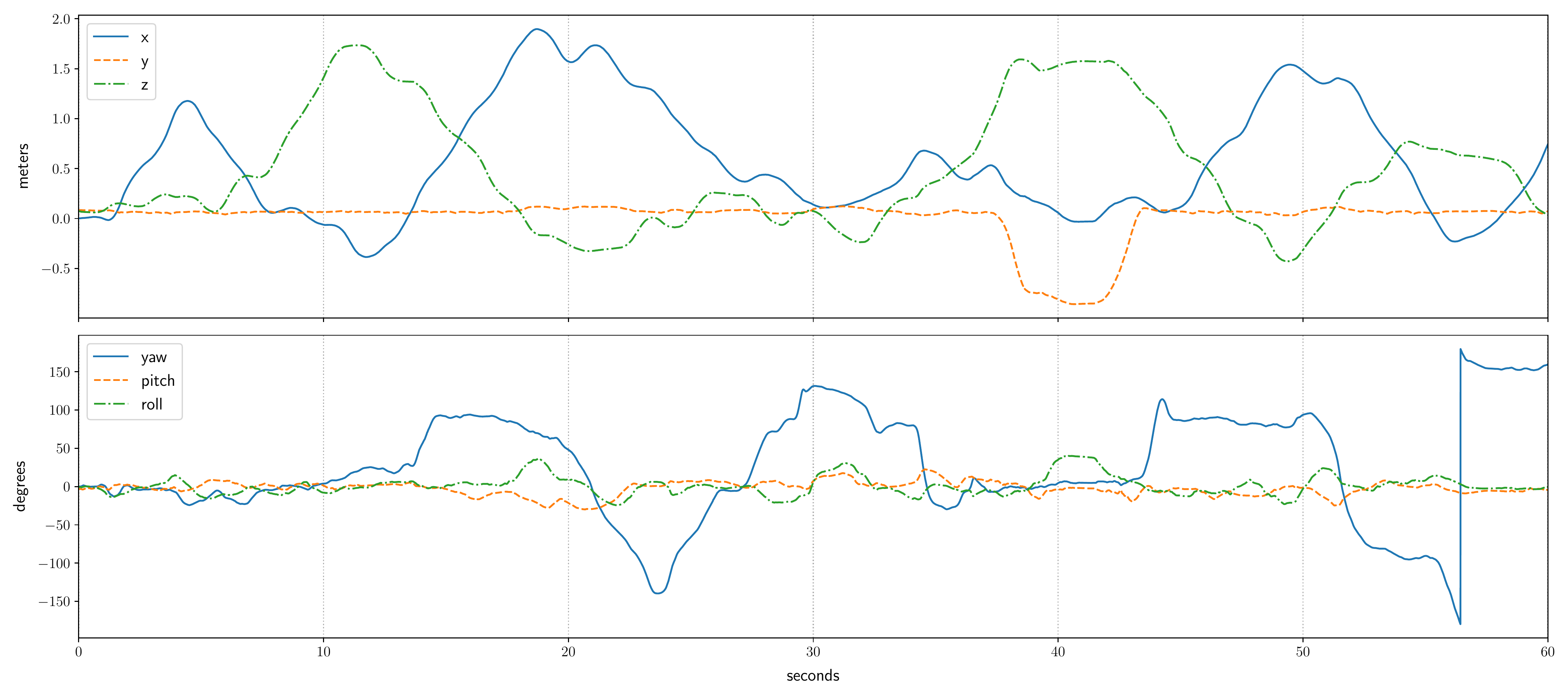}
	\caption[Sample trace]{One sample trace from our dataset collected using Microsoft HoloLens. Top: position, bottom: orientation (shown as Euler angles).}
	\label{fig:trace1}
\end{figure*}

Fig.~\ref{fig:system_kalman} shows the different steps of the Kalman filter-based predictor and presents interfaces for its integration into our volumetric streaming system.
After initialization, the state estimate from the previous step $\hat{\mathbf{x}}_{k-1}$ is propagated in time using the process model $\matr{F}$ and the time step of the filter $\Delta t=1/f_s$. 
Then, the obtained initial state estimate $\hat{\mathbf{x}}_{k}^\mhyphen$ is converted to a measurement estimate $\hat{\mathbf{z}}_k$ using the measurement model $\matr{H}$. 
Finally, the actual measurement $\mathbf{z}_k$ (sent over the network) is combined with $\hat{\mathbf{z}}_k$ to obtain the corrected state estimate $\hat{\mathbf{x}}_{k}$.
This completes one cycle of the Kalman filter operation. 
At the end of each cycle, the process model $\matr{F}$ is re-used to propagate the state estimate $\hat{\mathbf{x}}_{k}$ in time by the $\mathrm{LAT}=t_{\mathrm{p}}$ and obtain the predicted state $\hat{\mathbf{x}}_{k+N}$ at time $t_k+t_p$.
Finally, a corresponding view from the volumetric video is rendered based on the predicted user pose and transmitted to the client.

\section{Evaluation}
\label{sec:evaluation}
To evaluate the proposed predictors for different head movements and \glspl{lat}, we recorded user traces via Microsoft HoloLens and created a Python-based simulation framework that enables offline processing of the traces.
Below, we first discuss the experimental setup and the evaluation metrics, before presenting the obtained results and discussing the limitations of our approach.

\subsection{Experimental setup}
\label{sec:exp_setup}

\subsubsection*{Dataset}
We created a HoloLens application that overlays a virtual object (shown in Fig.~\ref{fig:skeleton}) on the real world and collected 14 head movement traces. 
For each session, the user was asked to freely move around and explore the object for a duration of \SI{60}{s}.
We recorded the position samples ($x$, $y$, $z$) and rotation samples in the form of quaternions ($q_x$, $q_y$, $q_z$, $q_w$) together with the corresponding timestamps. 
Since the raw sensor data we obtained from the HoloLens was unevenly sampled at \SI{60}{Hz} (i.e. different temporal distances between consecutive samples), we interpolated the data to obtain temporally equidistant samples. 
We upsampled the position data using linear interpolation and quaternions using \gls{slerp}~\cite{shoemake1985}. 
Thus, we obtained an evenly-sampled dataset with a sampling rate of \SI{200}{Hz}.
This resampling significantly simplifies the offline analysis of our traces and implementation of the predictors.\footnote{In an online setting, where an interpolation may not be feasible due to sequential arrival of the data points, a Kalman filter can be designed to handle varying time intervals between incoming samples~\cite{labbe2015}.}

\subsubsection*{Autoregression model settings}
We used one of our collected head motion traces (see Sec.~\ref{sec:exp_setup}) as training data and estimated the \gls{autoreg} model parameters from each time series ($x$, $y$, $z$, $q_w$, $q_x$, $q_y$, $q_z$) separately using the Python library \emph{statsmodels}~\cite{statsmodels}.
To select the best AutoReg model, we trained different models using three different training traces and selected the best-performing model.
Our models have an automatically determined lag order of 40 samples, i.e. they consider the past $40*5=200$~ms and predict the next sample using Eq.~\eqref{eq:autoreg}.

\subsection{Trace statistics}
\label{sec:trace_stats}

Fig.~\ref{fig:trace1} shows one of the traces in our dataset.
For visualization purposes, orientations are given as Euler angles (yaw, pitch, roll), although we perform the prediction in the quaternion domain (see Sec.~\ref{sec:orientation}).
Note that our framework uses a left-handed coordinate system where +y axis points up and +z axis lies in the viewing direction.

We can make two important observations based on the sample trace: firstly, the viewer rarely moves along the y-axis (except for the time period between \SIrange{38}{42}{s} during which the viewer probably sat down and stood back up), which is understandable since it requires more effort to crouch down and stand up.
Secondly, the orientation changes are typically due to yaw movements, whereas the magnitude of changes due to roll and pitch movements are much smaller.
Our observations are also confirmed by visual inspection of the other recorded traces.

\subsubsection*{Head movement velocity}
We analyzed the peak and mean head movement velocities by computing the first-order difference for all degrees of freedom ($x$, $y$, $z$, yaw, roll, pitch).
Since numerical differentiation using finite-differences is a noisy operation that amplifies any noise present in the data~\cite{chartrand2011}, we used a Savitzky-Golay filter~\cite{savitzky1964} to smooth the computed velocities.

\begin{figure}[ht]
	\includegraphics[width=\linewidth]{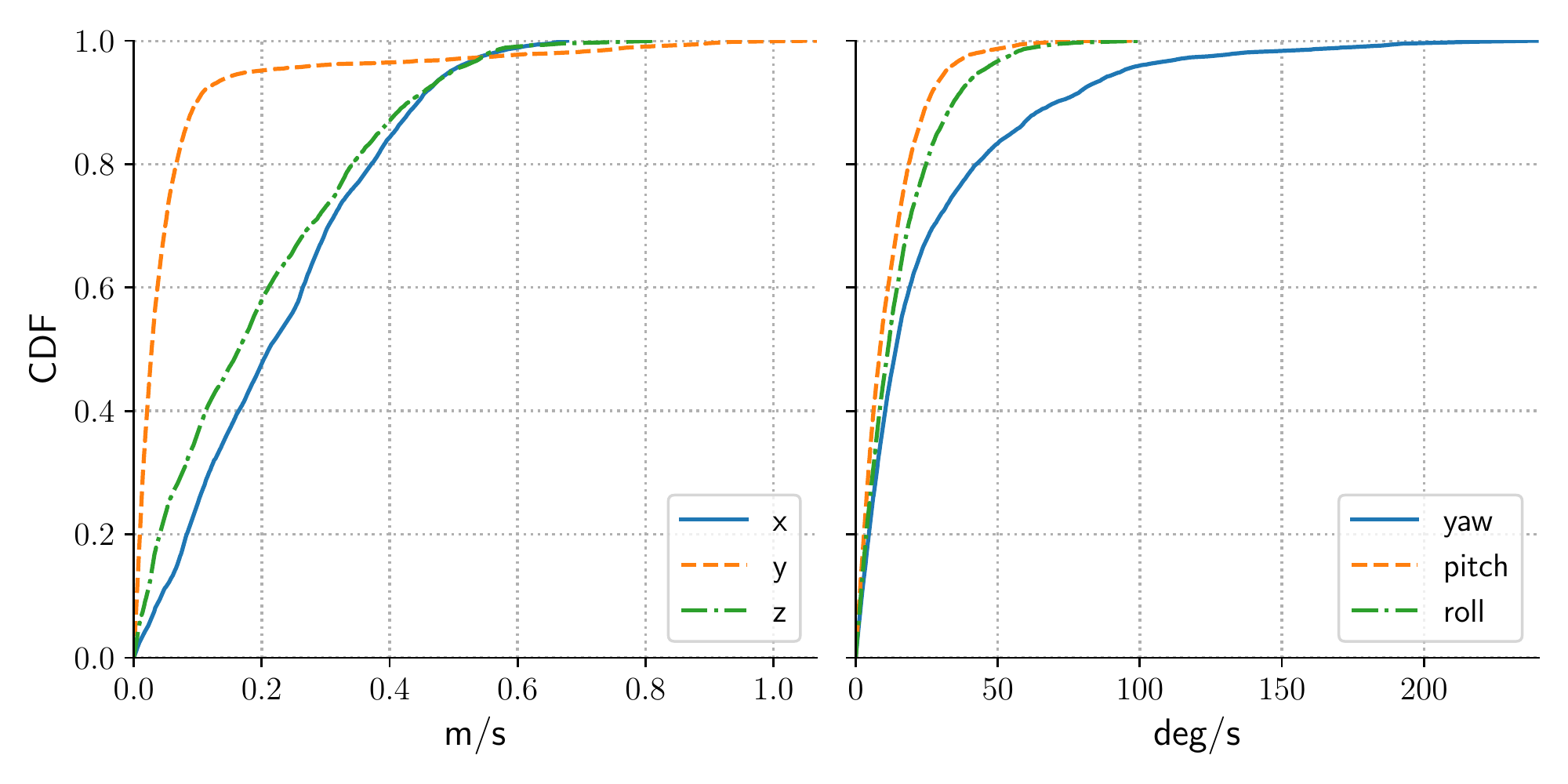}.
	\caption{\gls{cdf} of linear velocity (left) and angular velocity (right) for trace 1.}
	\label{fig:trace1_cdf}
\end{figure}

Fig.~\ref{fig:trace1_cdf} shows the \glspl{cdf} of the computed linear and angular velocities, respectively, for trace 1 (shown in Fig.~\ref{fig:trace1}).
We observe that the 95th percentile in $y$ dimension is located at \SI{0.2}{m/s}, whereas the 95th percentiles in $x$ and $z$ dimension are at around \SI{0.5}{m/s}.
We also observe that the pitch and roll velocities are mostly smaller than the yaw velocity, which can reach peak values around \SI{200}{deg/s}.

\begin{figure}[ht]
	\includegraphics[width=\linewidth]{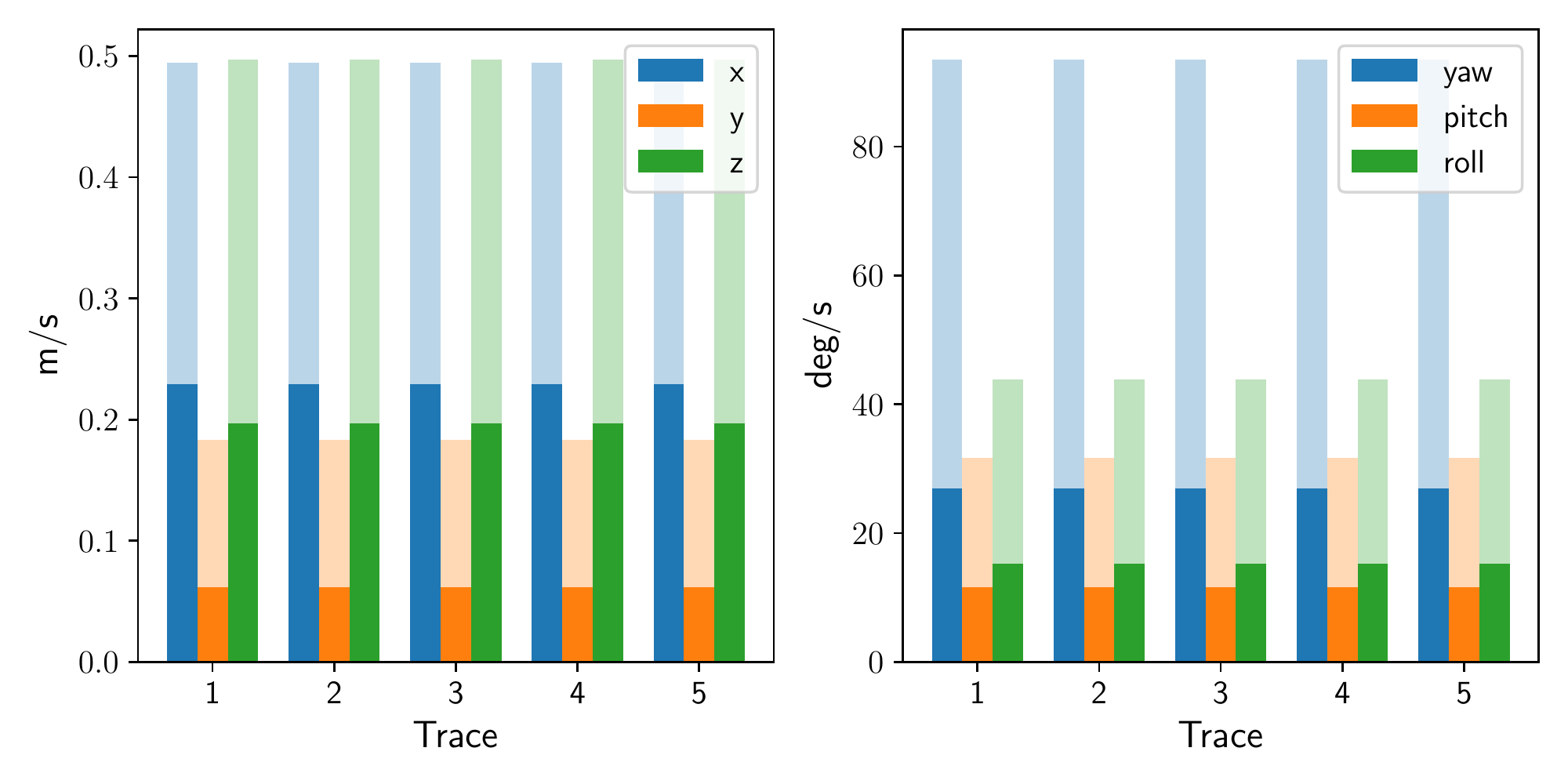}.
	\caption{Mean linear velocity (left) and mean angular velocity (right) for five traces. Lighter shades show the 95th percentile.}
	\label{fig:trace_velocity}
\end{figure}

Fig.~\ref{fig:trace_velocity} shows the mean linear and angular velocities as well as the 95th percentile ranges for five different traces.
In all traces, we observe that the linear velocities in $x$ and $z$ dimensions are greater than those in $y$. Similarly, the angular velocities in yaw dimension are greater than those in pitch and roll.
Deviations from the mean values can be significant in all dimensions, as observed by inspecting the 95th percentiles (lightly shaded in the figure). 

\begin{figure*}[t]
	\centering
	\includegraphics[width=\textwidth]{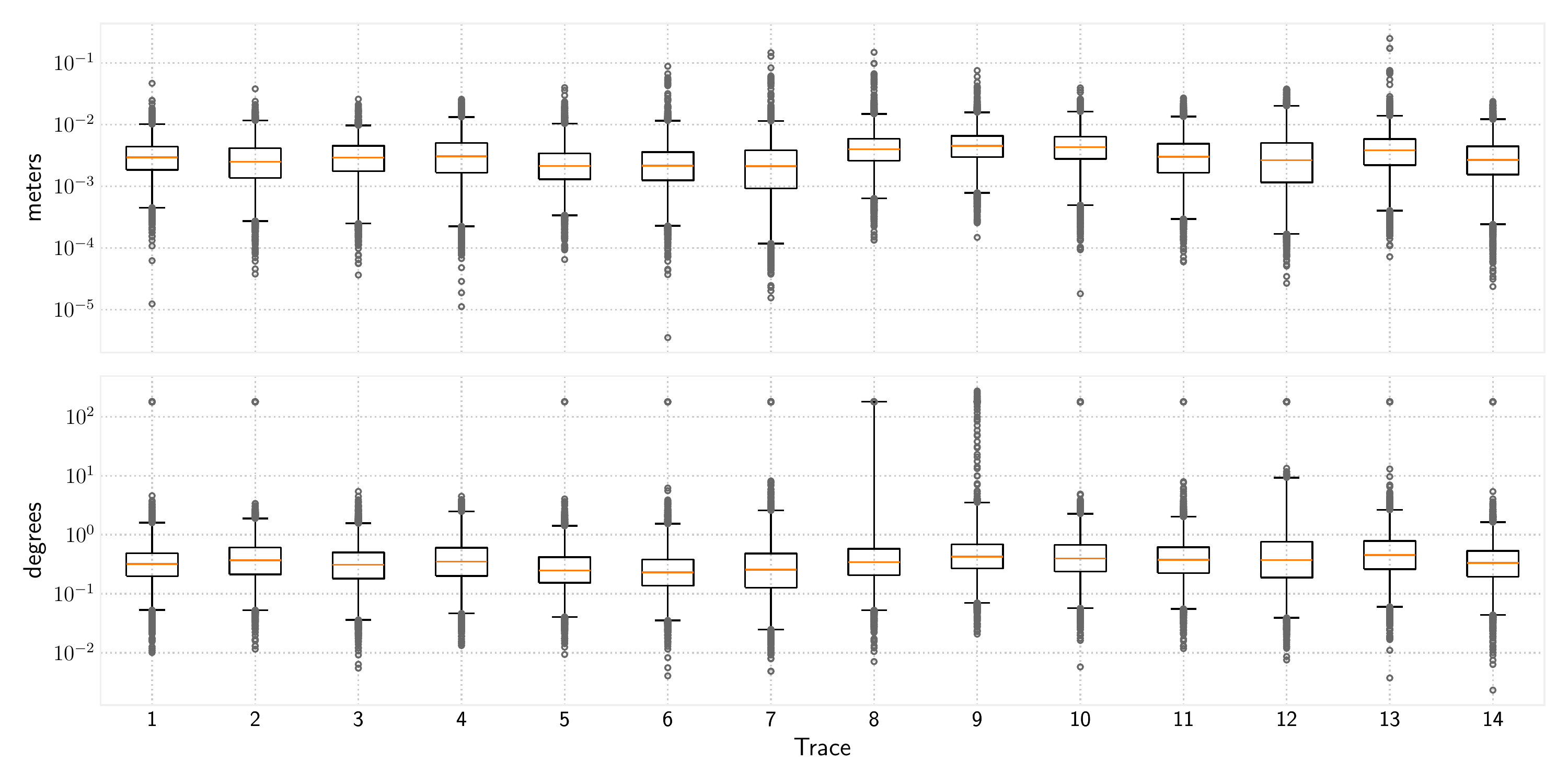}
	\caption{Per-trace distribution of the position errors (top) and angular errors (bottom) of the Kalman filter for a given LAT=60~ms. Whiskers represent the 1-99\% range; outliers are shown by circles. Each trace contains 12k data points. Note the logarithmic scale of the y-axes.}
	\label{fig:boxplot1-99}
\end{figure*}

\subsection{Evaluation metrics}
\label{sec:metrics}
For evaluation of the prediction methods, we employed two objective error metrics: position error and angular error. 
Position error is the Euclidean distance (in meters) between the actual and the predicted position. It is defined as
\begin{align}
d = \sqrt{(\hat{x}-x)^2 + (\hat{y}-y)^2 + (\hat{z}-z)^2}.
\end{align}
Angular error is the spherical distance (in degrees) between the actual and the predicted orientation. 
Let $\mathbf{q}$ be a measured and $\mathbf{\hat{q}}$ be a predicted unit quaternion.
Then, the spherical distance $\phi$ between the two orientations can be computed as follows \cite{sola2017}
\begin{align}
	\mathbf{r} &=\mathbf{q^*}\mathbf{\hat{q}} \\
	\phi &= 2\frac{180}{\pi}\arccos(r_w)
\end{align}
where $\mathbf{q^*}$ is the conjugate of the quaternion  $\mathbf{q}$ and $r_w$ is the scalar (real) part of the quaternion $\mathbf{r}$. 

After computing the position and angular error for each time point, we compute the \gls{mae} over a trace as 
\begin{align}
\mathrm{MAE(d)} = \frac{1}{N}\sum_{i=1}^{N}d_i,\;\;
\mathrm{MAE(\phi)} = \frac{1}{N}\sum_{i=1}^{N}\phi_i
\end{align}
where $d_i$ and $\phi_i$ are the position and angular errors at time $i$, respectively, and $N$ is the number of predicted samples in a trace.

\subsection{Prediction results}
\label{sec:results}
We evaluated the performance of our Kalman filter-based predictor for different \glspl{lat} \{20,40,60,80,100\}~ms. 
This range was chosen considering the measured \gls{m2p} latency of our cloud-based volumetric streaming system.
Running our server on an Amazon Web Services (AWS) instance, we measured an average \gls{m2p} latency around 60 ms with a network latency of 13.3 ms~\cite{anonym}.

First, we evaluate the performance of the Kalman filter for a fixed \gls{lat}, $t_p=60$~ms.
Fig.~\ref{fig:boxplot1-99} shows the distributions of the position and angular errors for each trace.
In the top plot, we observe that most of the position errors lie within the range \SIrange{0.1}{1}{cm} with few outliers exceeding \SI{10}{cm}.
Thus, the worst outliers are approximately one order of magnitude greater than the median of the position errors.
The bottom plot shows that the most angular errors have a magnitude of about \SIrange{0.1}{1}{deg}.
However, we notice that each trace contains a few angular errors around \SI{100}{deg} that would significantly degrade the accuracy of the rendered texture in a real system.
We believe that these outliers are caused by the inability of the standard Kalman filter to deal with the inherent properties of spherical distributions~\cite{kurz2016}.
In our case, those are the head orientations expressed as quaternions.
Although we applied a workaround to reduce their occurrence (see the discussion in Sec.~\ref{sec:limitations}), we could not yet find a way to eliminate all significant angular errors, as evident from the outliers in Fig.~\ref{fig:boxplot1-99}.

\begin{figure*}[ht]
	\centering
	\includegraphics[width=\textwidth]{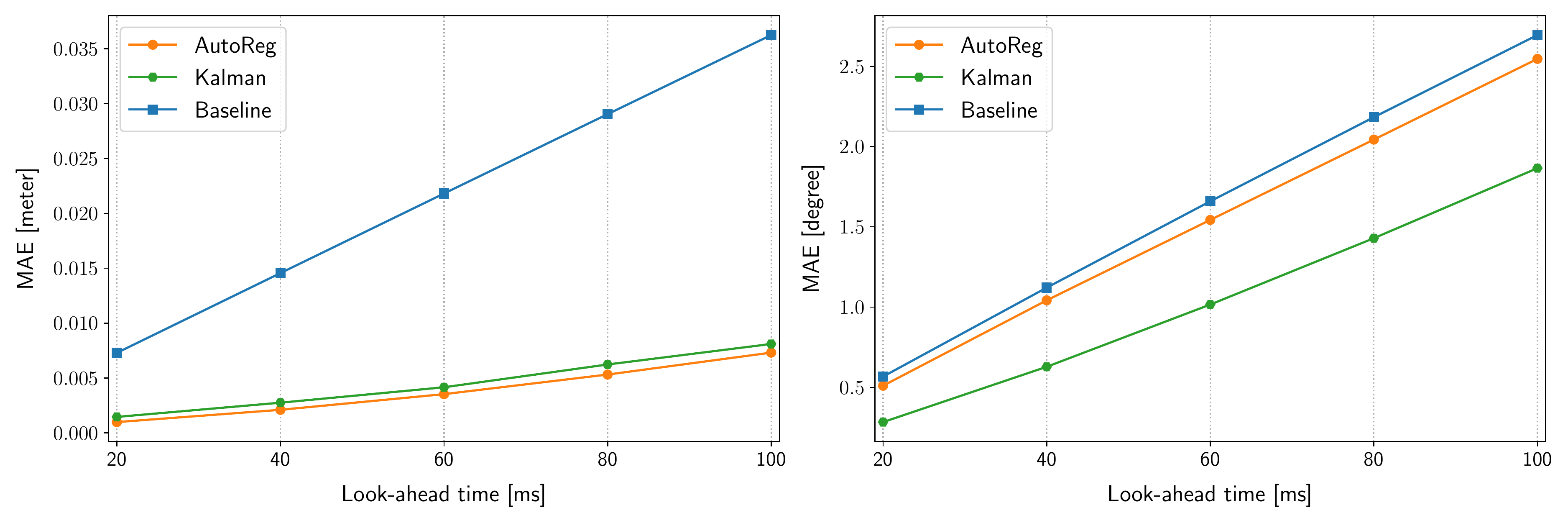}  
	\caption{Position error (left) and angular error (right) in terms of MAE. Averages over 14 traces are shown.}
	\label{fig:avg_mae}
\end{figure*}

Next, we compare the performance of the Kalman filter to the \gls{autoreg} and Baseline models and also aim to understand the effect of \gls{lat} on the prediction accuracy.
Fig.~\ref{fig:avg_mae} shows the mean $\mathrm{MAE(d)}$ and $\mathrm{MAE(\phi)}$ of the \gls{autoreg}, Kalman filter and the Baseline model for each \gls{lat}. 
The results are averaged over all traces.
We observe that both position error and angular error increase linearly with increasing \gls{lat}.
Both \gls{autoreg} and Kalman filter perform better than the Baseline in terms of position error showing that predicting translational movement is better than doing no prediction.
This is more evident for larger \gls{m2p} latencies (assumed equal to \glspl{lat} in our simulations) for which the performance of the Baseline deteriorates more quickly than both predictors.
However, the results for angular error show that \gls{autoreg} does not have a clear advantage over the Baseline. On the other hand, the Kalman filter decreases the angular errors on average by \SIrange{0.3}{0.9}{deg} depending on \gls{lat} compared to the Baseline.

\subsubsection*{Statistical significance}
To verify that the gain obtained by the Kalman filter over Baseline is statistically significant, we applied a two-sample T-test\footnote{A two-sample T-test checks the null hypothesis that two independent samples have identical average (expected) values~\cite{hogg2005}.} on the angular errors obtained by the Baseline and Kalman filter, for each trace and \gls{lat} combination.
Consequently, we reject the null hypothesis $\mathrm{H}_0$ of identical means of the angular error of the Baseline model and the Kalman filter-based prediction with $p<0.05$.

\subsection{Limitations}
\label{sec:limitations}
During the trace recordings, we observed that the HoloLens flips the sign of a quaternion, when its real (scalar) component $q_w$ approaches $\pm 0.5$.
Since a quaternion $\mathbf{q}$ and its negative $\mathbf{-q}$ correspond to the same orientation on the unit sphere~\cite{sola2017}, this behavior does not lead to a discontinuity in the rotation space.
However, the Kalman filter assumes that all latent and observed variables have a Gaussian distribution which cannot model the true topology of the spherical quantities~\cite{kurz2016}.
Therefore, these "jumps" in our data are detrimental to the performance of our predictor and cause large peak errors.

As a workaround, we compare a new measurement $\mathbf{z}_k$ with the previous one $\mathbf{z}_{k-1}$ at each iteration of the filter and reset the filter by re-initializing the state $\mathbf{x}$ and error covariance $\matr{P}$, whenever a sign flip is detected.
This workaround partially alleviates the peak errors observed after quaternion sign flips.
However, the filter requires a few iterations to output good predictions again which causes a few large outliers after a re-initialization.
In Sec.~\ref{sec:conclusion}, we identify some techniques that correctly handle spherical distributions and thus may reduce the observed peak errors.

\section{Conclusions and Future work}
\label{sec:conclusion}

This paper presented a Kalman filter-based framework for prediction of head motion in order to reduce the interaction (motion-to-photon) latency of a cloud-based volumetric streaming system.
We evaluated our approach using real head motion traces for different look-ahead times.
Our results show that the proposed approach can predict head orientation  \SIrange{0.3}{0.9}{deg} more accurately than the benchmark \gls{autoreg} model for a tested range of \glspl{lat} \SIrange{20}{100}{ms}.
Moreover, once its parameters are tuned, the Kalman filter is more robust to variations in data compared to the \gls{autoreg} model that has varying performance depending on the training data and needs training to learn its model coefficients. 

However, the presented approach exhibits several shortcomings that can be addressed in future research.
Particularly, our approach can be improved in terms of predicting spherical quantities.
In our future work, we will use recursive filters based on spherical distributions such as von-Mises-Fisher~\cite{kurz2016unscented} and Bingham distribution~\cite{kurz2016unscented} to predict head orientations more accurately.
These techniques rely on circular statistics and correctly handle the estimation in which the state is represented by a point on the unit sphere~\cite{kurz2016}.

Another promising direction is to take into account the content properties of the volumetric video (in addition to sensor measurements).
Such content-based techniques that take into account the visual saliency have recently been successfully applied for viewport prediction in 360-degree videos~\cite{aladagli2017, nguyen2018, ozcinar2019}.
We expect that extending these techniques to 6DoF volumetric videos can significantly improve the prediction accuracy.

Another open research area is the subjective evaluation of volumetric videos in AR/MR environments.
Our evaluation of prediction accuracy was performed using the objective metrics position error and angular error. 
However, although these metrics give a good idea about the relative performance of different predictors, the effect on user experience in a real system cannot directly be inferred from our results. 
Moreover, AR/MR headsets like Microsoft HoloLens use post-rendering updates known as late-stage reprojection ~\cite{hololens_lsr} or time-warping~\cite{timewarp} to account for any slight head movement since the last head pose prediction that may cause "judder" effects, i.e. unstable overlaid virtual objects.
Through image warping, such corrections can compensate for prediction errors in \gls{6dof} motion~\cite{mark1997}.
Therefore, subjective tests are required to understand the effect of prediction together with post-rendering correction on the user perception in an \gls{ar}/\gls{mr} environment.
In this regard, we are currently investigating the effect of latency and mispredictions on the subjective experience of the viewers.


\bibliographystyle{ACM-Reference-Format}
\bibliography{acmmm20.bib}


\end{document}